\documentclass[sigconf]{acmart}

\AtBeginDocument{%
  \providecommand\BibTeX{{%
    \normalfont B\kern-0.5em{\scshape i\kern-0.25em b}\kern-0.8em\TeX}}}

\copyrightyear{2022}
\acmYear{2022}
\setcopyright{acmcopyright}\acmConference[ICSE '22 Companion]{44th International Conference on Software Engineering Companion}{May 21--29, 2022}{Pittsburgh, PA, USA}
\acmBooktitle{44th International Conference on Software Engineering Companion (ICSE '22 Companion), May 21--29, 2022, Pittsburgh, PA, USA}
\acmPrice{15.00}
\acmDOI{10.1145/3510454.3516848}
\acmISBN{978-1-4503-9223-5/22/05}

\usepackage{soul}
\usepackage{float}
\usepackage{multirow}
\usepackage{epstopdf}
\usepackage{hyperref}
\usepackage{listings}
\usepackage{fancybox}
\usepackage{graphicx}
\usepackage{subcaption}
\usepackage{paralist}
\usepackage{ragged2e}
\usepackage{enumitem}

\usepackage{amsmath,bm}

\usepackage[framemethod=TikZ]{mdframed}
\usepackage{tcolorbox}
\makeatletter
\newcommand{\mybox}[1]{%
  \setbox0=\hbox{#1}%
  \setlength{\@tempdima}{\dimexpr\wd0+13pt}%
  \begin{tcolorbox}[boxrule=0.5pt, colback=white, arc=4pt,
      left=6pt,right=6pt,top=6pt,bottom=6pt,boxsep=0pt]
    #1
  \end{tcolorbox}
}
\usepackage{tikz}
\usepackage{pgfplots}
\usetikzlibrary{pgfplots.statistics,calc}
\usepackage{color}
\usepackage{xcolor}
\usepackage{array}
\usepackage{amsmath}
\usepackage{centernot}
\usepackage{xspace}
\usepackage{url}
\usepackage{verbatim}
\usepackage{wrapfig}
\usepackage{tabularx}
\usepackage{bbding}
\usepackage{pifont}
\usepackage{wasysym}
\usepackage{amssymb}

\newcommand{\tool}{\texttt{NaviDroid}}

\makeatletter  
\newif\if@restonecol  
\makeatother  
\usepackage[linesnumbered,ruled,vlined]{algorithm2e}
\usepackage{algpseudocode}  
\usepackage{amsmath}  

\begin{document}

\title{NaviDroid: A Tool for Guiding Manual Android Testing via \\ Hint Moves}

\author{Zhe Liu$^{1,3}$, Chunyang Chen$^2$, Junjie Wang$^{1,2,3,*}$, Yuhui Su$^{1,3}$, Qing Wang$^{1,2,3,*}$}
\affiliation{
  \position{$^1$Laboratory for Internet Software Technologies, $^2$State Key Laboratory of Computer Sciences, }
  \department{Institute of Software Chinese Academy of Sciences, Beijing, China; \\
  $^3$University of Chinese Academy of Sciences, Beijing, China; $^*$Corresponding author\\
  $^4$Monash University, Melbourne, Australia;
  }
}
\email{liuzhe181@mails.ucas.edu.cn, Chunyang.chen@monash.edu, junjie@iscas.ac.cn, wq@iscas.ac.cn}


\begin{abstract}
Manual testing, as a complement to automated GUI testing, is the last line of defense for app quality especially in spotting usability and accessibility issues.
However, the repeated actions and easy missing of some functionalities make manual testing time-consuming, labor-extensive and inefficient.
Inspired by the game candy crush with flashy candies as hint moves for players, we develop a tool named {\tool} for navigating human testers via highlighted next operations for more effective and efficient testing.
Within {\tool}, it constructs an enriched state transition graph (STG) with the trigger actions as the edges for two involved states. Based on the STG, {\tool} utilizes the dynamic programming algorithm to plan the exploration path, and augment the run-time GUI with visualized hint moves for testers to quickly explore untested states and avoid duplication.
The automated experiments demonstrate the high coverage and efficient path planning of {\tool}.
A user study further confirms its usefulness in the participants covering more states and activities, detecting more bugs within less time compared with the control group. 

\textit{NaviDroid demo video: \url{https://youtu.be/lShFyg_nTA0}.}

\end{abstract}

\keywords{GUI testing, Android App, State Transition Graph, Human testing}

\maketitle

\section{Introduction}
\label{sec_introduction}
With the development of mobile devices, mobile apps are indispensable for people's daily life in accessing the world.
The importance of mobile apps makes it vital for the development team to carry out a thorough testing for ensuring the quality of mobile apps~\cite{feng2021auto}.
However, mobile apps are event-centric programs with rich graphical user interfaces (GUIs)~\cite{DBLP:conf/sigsoft/XieFXCC20,zhao2020seenomaly,zhao2021guigan}, and interact with complex environments. 
To ensure the app quality, there are mainly two kinds of GUI testing i.e., automated GUI testing and manual GUI testing.
The automated GUI testing studies for mobile apps include model-based, probability-based and deep learning-based approaches~\cite{Wang2021Vet,pan2020reinforcement,li2019humanoid,yang2018static}. 
Albeit its convenience and scalability, such automated testing may not have a high activity coverage, especially for those functionalities which can only be reached by complicated inputs or actions~\cite{alshayban2020accessibility,liu2020owl}. Furthermore, the usability and accessibility bugs (e.g., color schema, font size, interaction)~\cite{chen2021accessible,yang2021uis,yang2021don,liu2022Nighthawk} are difficult to reveal by automated GUI testing.

\begin{figure}[htb]
\centering
\vspace{-0.05in}
\includegraphics[width=8.5cm]{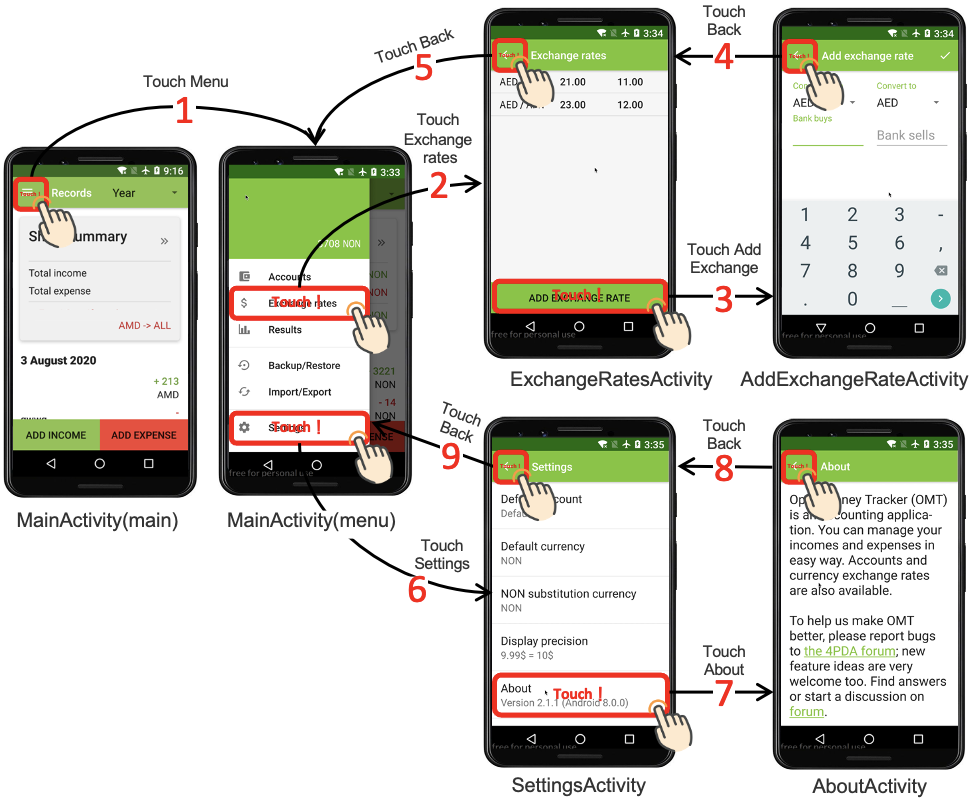}
\vspace{-0.1in}
\caption{Example {\tool} usage scenario.}
\label{fig:NaviDroid_example}
\vspace{-0.1in}
\end{figure}

\begin{figure*}[t]
\centering
\vspace{0.1in}
\includegraphics[width=17.8cm]{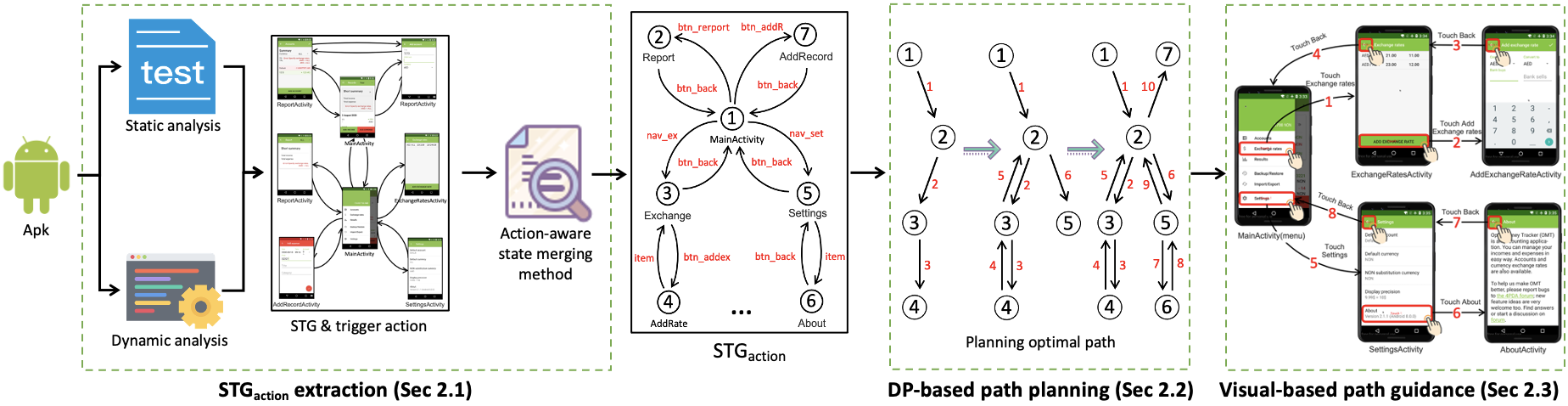}
\vspace{-0.1in}
\caption{Overview of {\tool}.}
\label{fig:overview}
\vspace{-0.1in}
\end{figure*}

Therefore, in addition to automated testing, companies also adopt manual testing as the last line of defense~\cite{linares2017developers}.
Research also showed that due to the usability and learning curve of automated tools, manual testing is preferred by many software developers~\cite{linares2017developers,wang2021context,DBLP:journals/tse/WangWCMCXW21,wang2019iSENSE}.
Compared with automated testing, human testers can discover more diverse and complicated bugs, especially those related to user experience.
However, manual GUI testing also has the following challenges.
First, it is time-consuming which requires many testers to manually explore each UI page of the app, and they may execute repeated actions. 
Second, the performance of manual testing is unstable as it highly depends on the testers' capability and experience, and testers may miss some minor functionalities. To leverage the pros of both testing techniques, we propose a new hybrid approach, {\tool}, to assist manual GUI testing based on the insights from the automated GUI testing.

{\tool} is a user-friendly app. Testers can upload the Android APK and get real-time guidance.
Inspired by the automated testing, {\tool} distills the prior knowledge of one app including all states and state relationships.
During manual GUI testing, {\tool} will trace testers' testing steps and help navigate or remind testers with unexplored pages by explicit visual annotations (e.g., red bounding box) in the run-time page as seen in Fig \ref{fig:NaviDroid_example}.
That process is similar to the flashing candies (hint/suggested moves) when a player hesitates to make a move in playing Candy Crush (a popular free-to-play match-three puzzle video game)~\cite{Candy2020}.
According to our observation, that suggested hint move is particularly useful when the trigger components are small or poorly designed/developed. 
It can help human testers avoid missing some functionalities or making repeated exploration steps. 

The contributions of this paper are as follows:

\begin{itemize}
\item We implement a static and dynamic based STG extraction method and a DP based path planning method to guide the path exploration in covering all the states with few repeated exploration steps.

\item We develop a fully automated app {\tool}. Users only need to upload an Apk file, and {\tool} will automatically guide users to test the app. We release the implementation of {\tool} on GitHub.

\item An empirical study among professionals proves the usefulness of {\tool} in assisting manual GUI testing and finding practical bugs.

\end{itemize}

\section{Approach}
\label{sec_approach}
This paper proposes {\tool} to navigate testers in exploring apps to avoid missing functionalities or making repeated exploration steps.
Figure \ref{fig:overview} presents the overview of {\tool}, which consists of three main components.
\textbf{\textit{First}}, given the app Android package (Apk), $STG_{action}$ extraction component combines both static analysis and dynamic exploration to extract the state transition graph (STG) and its trigger actions between states (Section \ref{subsec_approach_ATG_extraction}). 
We also design a context-aware state merging method to merge near-duplicate states by considering the current state and the adjacent states. 
\textbf{\textit{Second}}, based on the extracted $STG_{action}$, the DP-based path planning component plans the exploration path which aims at covering all the states of the app with a few repeated exploration steps (Section \ref{subsec_approach_Path}).
\textbf{\textit{Third}}, with the planned path, the visual-based path guidance component utilizes the visual augmentation technology to guide users' testing (Section \ref{subsec_approach_Visual}). 
For more details please refer to our full paper~\cite{liu2022CHI}.

\subsection{$STG_{action}$ Extraction}
\label{subsec_approach_ATG_extraction}

We first extract the state transition graph (STG), and then enrich it with trigger actions between the states to construct $STG_{action}$, which serves as the basis in guiding the users exploring the app.
In our method, $STG_{action}$ is defined as a graph $G <N, E>$ with node $N \in state$ and edge $E \in action$.
\textbf{State}: Inspired by app GUI testing~\cite{pan2020reinforcement}, we regard each unique UI page as one $state$ and represent it by i.e., represented by UI components hierarchy tree.
Each $activity$ may have multiple $state$, that is, $N \in state \in activity$.
\textbf{Action}: Action is the trigger that results in $state$ transition, which can be expressed as $E = ID, \ E \in action$. 

\textbf{Static STG Extraction and Trigger Action Detection.} 
In an Android application, activities can be started by invoking.
For example, the $StartActivity(intent)$ is an inter-component communication (ICC) call, passing an intent that describes the activity to be launched~\cite{zhang2018launch,yanmultiple}.
In detail, the target activity of ICC call is determined by querying the pointed-to values in the fields of an intent object. 
By matching the parameter in $intent()$ method with the parameter in \textit{AndroidManifest.xml} file, we obtain the transition between activities and build the initial $STG_{action}$.

\textbf{Dynamic STG Extraction and Trigger Action Detection.}
Some states and trigger actions, especially those in dynamic or mixed layout (such as dynamic rendering menu), are difficult to be obtained by static analysis.
Instead, it is easy to be captured with dynamic GUI rendering.
In detail, by leveraging the idea of dynamic app GUI testing~\cite{Wang2021Vet}, we adopt an app explorer~\cite{li2017droidbot} to automatically explore the pages within an application through interacting with apps using random actions e.g., clicking, scrolling, and filling in text. 
During the exploration, we record both $state$ and trigger $action$ between states. 
We then combine the $STG_{action}$ extracted from static analysis and dynamic exploration into one graph.

\textbf{Context-aware State Merging}
Through static and dynamic analysis, we get $STG_{action}$ which is composed of a large number of $states$ and $actions$, in which some of them are duplicates~\cite{li2017droidbot,Wang2021Vet}.
Given $STG_{action}$, we first merge the states with the same GUI run-time hierarchy without considering detailed content (e.g., text or image) which may change dramatically.
After that, we further merge states with similar GUI hierarchy by checking whether their $n-1$ $state$ (i.e., the previous state which transits to the current state) and $n+1$ $state$ are similar.

\subsection{DP-based Path Planning}
\label{subsec_approach_Path}

With the $STG_{action}$ obtained in the previous section, we need to plan a path that can cover all the nodes (i.e., states) with a few repeated steps, so as to serve as the basis for the follow-up testing guidance.
To achieve this, we use a dynamic programming algorithm to derive and plan the shortest path. 

\textbf{Formalization of Path Planning.}
We formulate the path planning as a dynamic programming problem, and represent it by a 4-tuple: $\bm{< G, d, V, DP>}$.
$\bm{G}$: \textbf{Graph}. The $STG_{action}$ ($\bm{G <N, E>}$) obtained in the previous section, where $\bm{N}$ is the set of nodes \textit{(i.e., states)}, and $\bm{E}$ is the set of edges \textit{(i.e., triggered events)}. 
$\bm{d}$: \textbf{Distance}. $\bm{d_{ij}}$ is the shortest distance between state $i$ and $j$. 
$\bm{V}$: \textbf{Visit status}. $\bm{V}$ is the visit status of the current node, represented by binary numbers. 0 is not visited and 1 is visited.
$\bm{DP}$: \textbf{Dynamic programming}. $\bm{DP_{jV}}$ is the shortest distance from the current state $i$ to state $j$ in visit status $\bm{V}$. Since $\bm{V}$ is a binary number, $\bm{DP_{i(V\land(1 \ll (j-1))}}$ is the distance of reaching state $i$ without accessing other states.

Under the above formalization, to solve the path planning problem is to optimize the following two equations:
$$\left\{
\begin{array}{l} 
d_{ij} = min(d_{ij}, d_{ik} + d_{kj})\\
DP_{jV} = min(DP_{jV}, DP_{i(V\land(1 \ll (j-1))} + d_{ij})
\end{array}
\right.
$$

\textbf{Planning Strategies.}
The general idea of dynamic programming algorithm is to use multi-stage optimal decision-making, where each decision depends on the current visit status, and then cause the visit status to transfer.
In detail, the algorithm first traverses the graph $STG_{action}$ to obtain the set of nodes $N$ and edges $E$. 
It then employs Floyd algorithm to calculate the shortest path between each pair of nodes $d[i][j]$. 
It maintains a buffer to store the visit status $V$, and gives priority to the nodes that have not been visited. 
Suppose the exploration is currently at node $i$, the algorithm will judge whether the visit status $DP[j][V]$ of node $j$ is visited; if not, it finds the shortest path $d[i][j]$ between node $i$ and node $j$, and update the node visit status $VisitStatus$.
After all nodes in the graph are visited, the algorithm can recommend the planned path for testers to explore.

\subsection{Visual-based Path Guidance}
\label{subsec_approach_Visual}

We further implement the planned path into {\tool} for guiding testers in testing mobile apps.
It can suggest the next operation step by step in the user interface to help the testers cover the unexplored pages and reduce the replication explorations. 
Specifically, we adopt the Android floating window ~\cite{chen2020improving} for visualizing the hint moves.
As the Android interface drawing is realized through the services of $WindowManager$, which can add the floating window control to the screen through the $AddView()$ method.
The system runs the floating window service in the backend, and sets the size and coordinates of the floating window to make it in the same position and suitable size and floating on the component, so as to guide the testers to explore the app's interface.

\section{Tool Implementation And Usage}
\label{sec_implementation}


\subsection{Implementation}
\label{subsec_Web}
{\tool} can automatically extract the STG of the app and guide testers to test the app according to the dynamically planned path. 
Specifically, we augment the run-time GUI with visual hint moves.
{\tool} uses the Android debug bridge (adb) command to start the application that testers need to test.
During the tester's exploration, {\tool} obtains the run-time information of the current state (interface) including the state information and existing components within the current page on the backend. 
Given the state information of the current page, {\tool} searches the obtained STG on the fly, finds the next state on the planned path, and highlights the corresponding actions which can trigger that state in the Android GUI page. 

Specifically, the {\tool} mainly provides the following three modules: extracting the STG automatically and display, planning the exploration path based on the DP algorithm, guiding the testers in real time.

\textbf{Extracting the STG automatically and display:} Testers can upload the Apk of the application to {\tool}. It will combine both static analysis and dynamic exploration to extract the STG of the app. The {\tool} then automatically generates the \textit{STG.html} to display the extracted STG.

\textbf{Planning the exploration path based on the DP algorithm:} According to the STG extracted from the previous module, the {\tool} uses the dynamic programming algorithm to plan the test path in backend. Note that if the tester does not follow the path recommended by our approach in the process of exploration, we would record the state when he/she changes the path. Then according to the path that has been explored, our {\tool} will recalculate the path by running the DP algorithm and take the current state as the starting point.

\begin{figure}[htb]
\vspace{-0.05in}
\centering
\includegraphics[width=8.3cm]{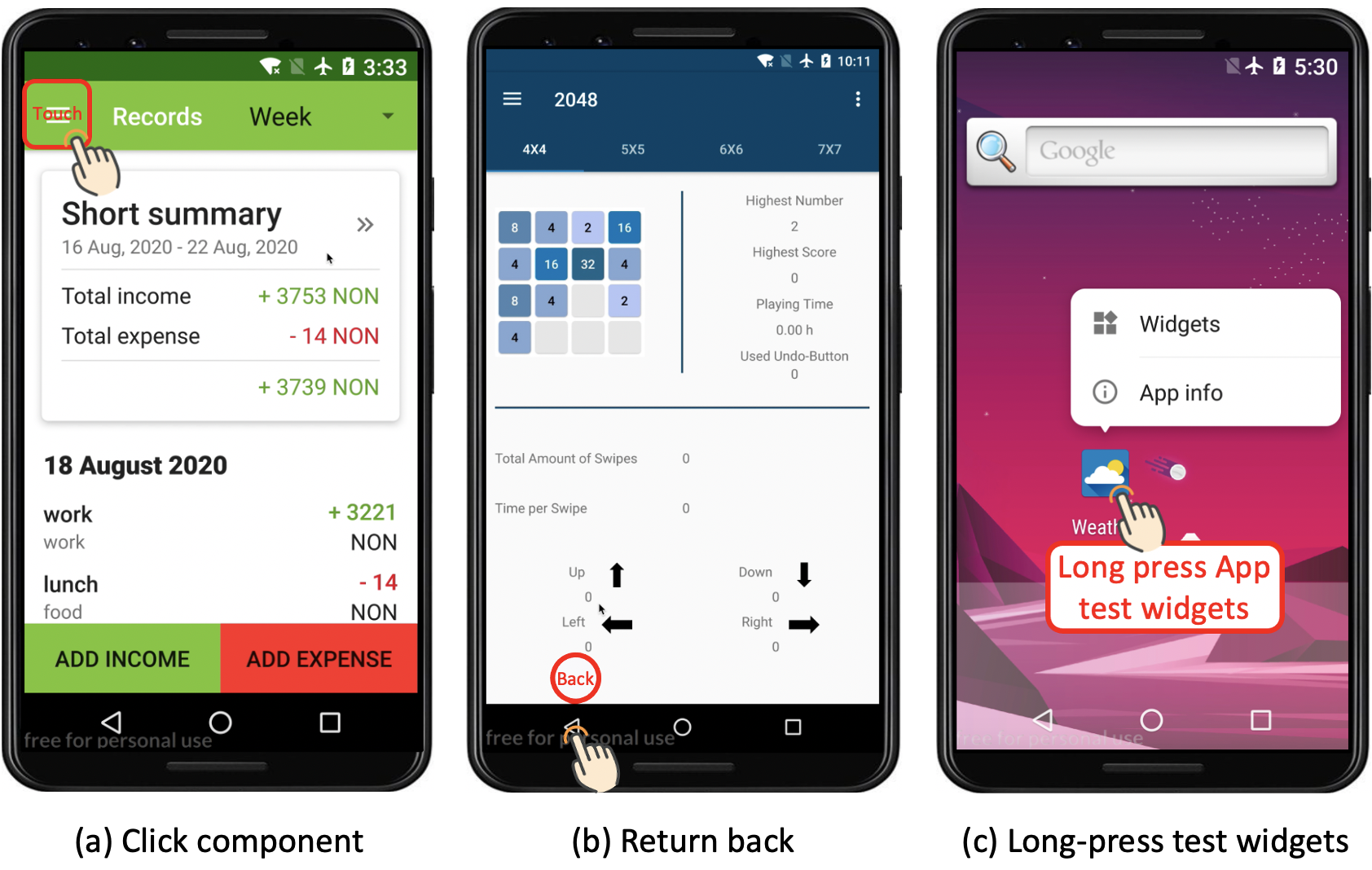}
\vspace{-0.1in}
\caption{Illustration of our {\tool}.}

\label{fig:Illustration}
\vspace{-0.1in}
\end{figure}

\textbf{Guiding the testers in real time:} Figure \ref{fig:Illustration} gives the example of the hint moves in real-time. The {\tool} runs the service in the backend, and sets the size and coordinates of the floating window (hint moves) to make it in the same position and floating on the component, so as to guide the testers to explore the app interface. we implement three types of trigger actions to provide a more friendly interactive experience, including:

\begin{itemize}[leftmargin=*]
\item \textit{\textbf{Clicking a component:}} In Figure \ref{fig:Illustration} (a). For a button, a navigation bar or a fragment bar, the tester is suggested to click the UI component.

\item \textit{\textbf{Returning back:}} In Figure \ref{fig:Illustration} (b). If there is no back button in the current interface and the id is `touch\_back', {\tool} will suggest the tester with ``back'' action above the back key, otherwise, the tool will directly highlight the back button.

\item \textit{\textbf{Long-pressing test widgets:}} In Figure \ref{fig:Illustration} (c). If ``appwidget activity'' exists in activity, 
for the ``appwidget activity'', the tester will be suggested to operate ``long press app test widgets'' through the floating window when exploring the app. 
\end{itemize}

\subsection{Usage Scenarios}
\label{subsec_Usage}

We present several examples to illustrate how testers would interact with {\tool}. In some cases, when facing an unfamiliar app, testers usually know the function of the app by random clicking. However, testers may execute repeated actions during the exploration. At the beginning of the test, tester can directly input the Apk of the app to our {\tool}. The {\tool} will automatically extract the $STG_{action}$ and guide the user to explore the app according to the $STG_{action}$. It can help testers understand the application in the shortest time.

During the test process, in order to improve the activity coverage of manual testing, the {\tool} dynamically plans the testing path according to the tested interface of the tester. Specifically, the {\tool} uses adb to detect the user's operation in the backend in real-time and records the access times of each state (interface). When the {\tool} detects that the developer has not operated for more than 5 seconds, it will take the current state as the starting point and use the dynamic programming algorithm to re-plan the path. Because the new path considers the state that the user has tested, it can avoid repetition and guide the user to test more states. If the tester does not follow the path planed by {\tool}, it will dynamically plan a new path according to the current state explored by the tester.


\section{Evaluation}
\label{sec_evaluation}

\subsection{Effectiveness Measurement}
\label{sec_effectiveness}

{\tool} consists of STG extraction algorithm, dynamic programming algorithm and visual guidance. Therefore, we first evaluate the algorithm performance of the {\tool} through an automated method.
We evaluate the effectiveness of {\tool} from the points view of $STG_{action}$ extraction component and DP-based path planning component respectively. 

Given the effectiveness of our {\tool} for $STG_{action}$ extraction and path planning, we evaluate the {\tool} on 85 open-source apps from F-Droid.
This part is also published in our previous work~\cite{liu2022CHI} and we mainly use evaluation metrics of state coverage and exploration steps.

\begin{figure}[htb]
\centering
\vspace{0.1in}
\includegraphics[width=8.4cm]{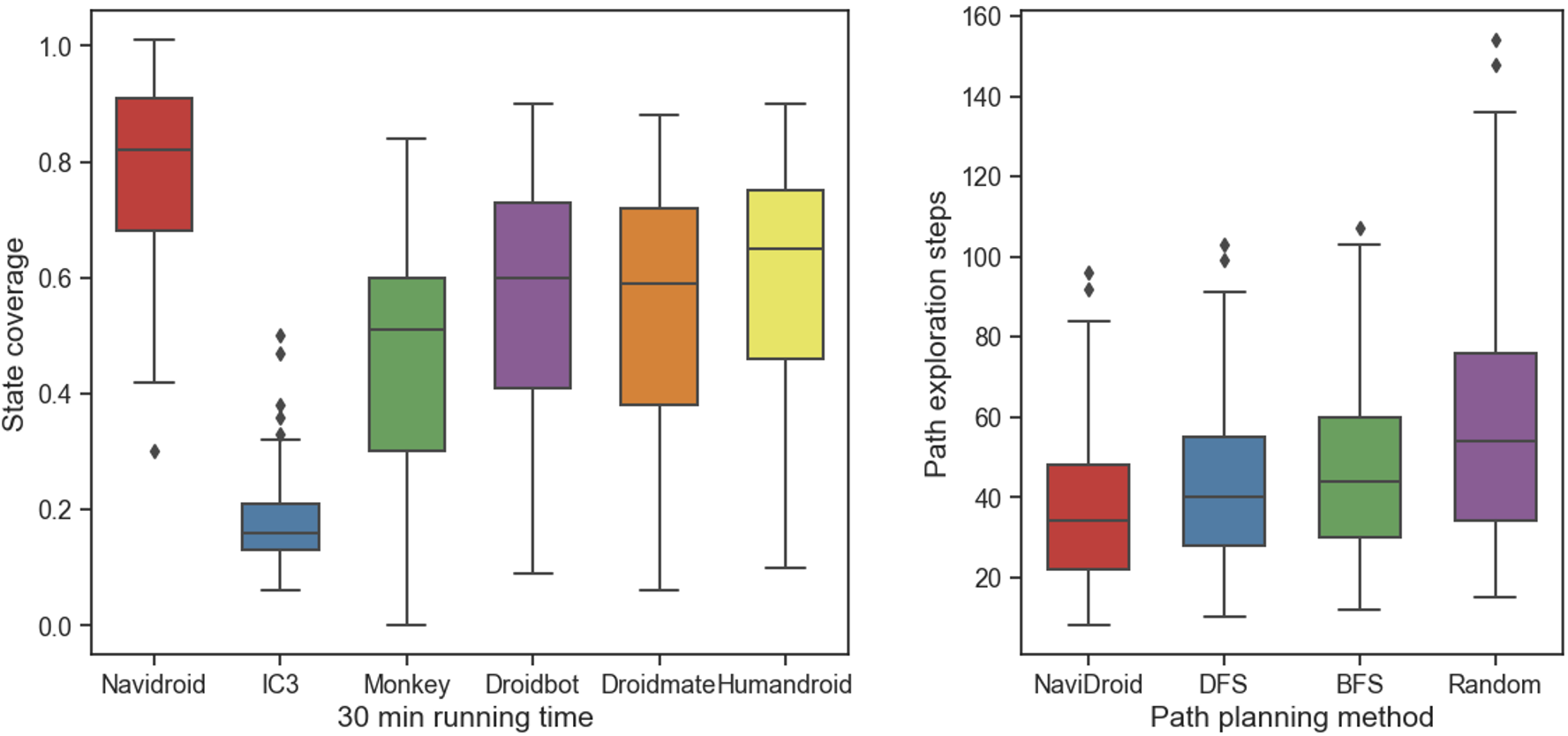}
\vspace{-0.1in}
\caption{Result of effectiveness evaluation.}

\label{fig:effectiveness}
\vspace{-0.1in}
\end{figure}

Figure \ref{fig:effectiveness} shows the performance comparison with the baselines. 
Results show that {\tool} can achieve 81\% median state coverage with the extracted $STG_{action}$, outperforming five commonly-used and state-of-the-art baselines. 
It also saves 20\% to 42\% exploration steps compared with the three commonly-used baselines. 

\subsection{Usefulness Measurement}
\label{sec_usefulness}

We further carry out a user study to evaluate its usefulness in assisting manual GUI testing, with 20 apps from F-droid. 
We recruit 32 testers to participate in the experiment from a crowdtesting platform TestIn. The experimental group (P1-16) who test the mobile apps guided by our {\tool}, and the control group (P17-32) who conduct the testing without any assistant.
This part is also published in our previous work~\cite{liu2022CHI}.
Results show that, the participants with {\tool} cover \textbf{62\% more states} and \textbf{61\% more activities}, \textbf{detect 146\% more bugs} within 3\textbf{3\% less time}, compared with those without {\tool}.
This confirms the usefulness of {\tool} in avoiding missing functionalities and making repeated exploration steps, and helping detecting bugs during manual testing.

\begin{figure}[htb]
\centering
\vspace{-0.05in}
\includegraphics[width=8.4cm]{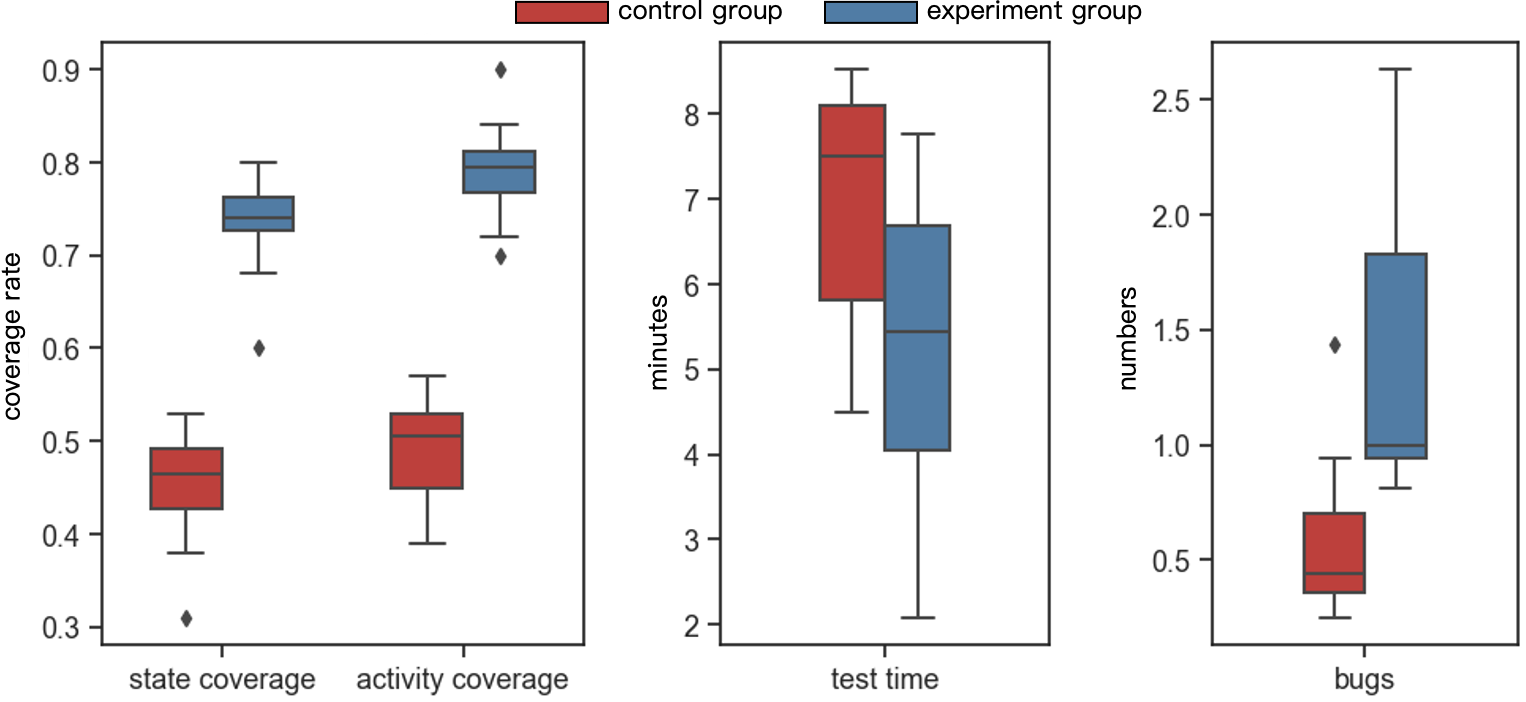}
\vspace{-0.1in}
\caption{Result of usefulness evaluation.}
\label{fig:usefulness}
\vspace{-0.1in}
\end{figure}

Regarding the user experience of {\tool}, we create an online survey on 30 professional testers and researchers, whom major in computer science or mobile testing. We ask them to use and feed back the usefulness of {\tool}, as well as its potential and scalability. In the end, they fill in the System Usability Scale (SUS) questionnaire (5-point Likert scale, e.g., 5 (strongly agree)). 

\begin{figure}[htb]
\centering
\vspace{0.1in}
\includegraphics[width=8.4cm]{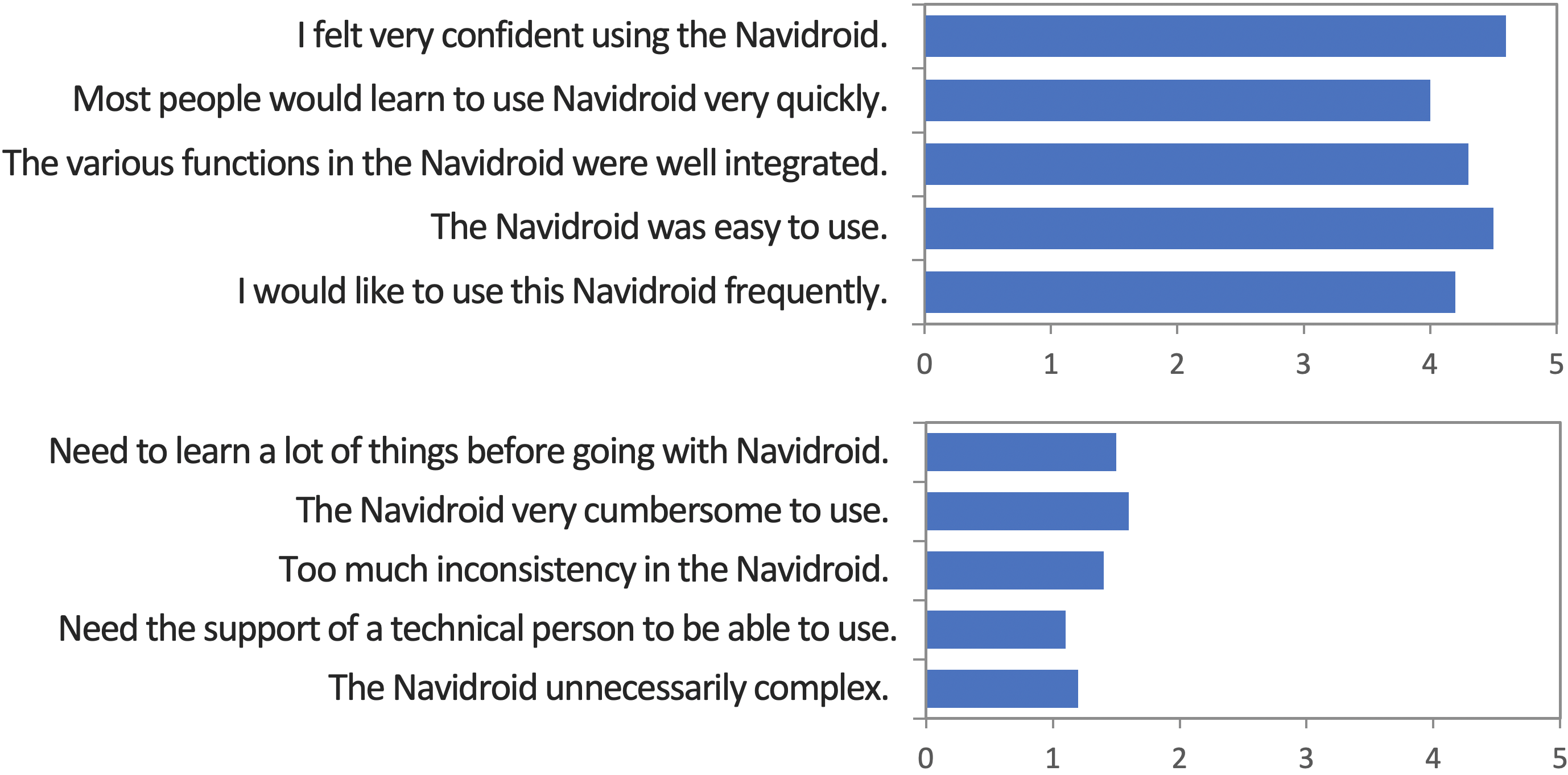}
\vspace{-0.1in}
\caption{Average score of SUS results.}
\label{fig:usefulness}
\vspace{-0.15in}
\end{figure}

Figure \ref{fig:usefulness} summarizes the participants’ ratings of the 10 system design and usability questions in the SUS. The upper half of Figure \ref{fig:usefulness} shows that they agree the features of the {\tool} are well-devised. The lower half of figure \ref{fig:usefulness} further confirms its simplicity and consistency. Furthermore, its average helpfulness for the tasks is 4.51, which indicates that they appreciate the help of {\tool}.
All of them appreciate that the hint moves can help guide them in exploring the inconspicuous UI pages, increasing the hit rate of potential bugs.
For example, ``{\tool} is very helpful for us to discover new pages and functions. It effectively avoids repeated operations.''.
Participants express they like our interaction design such as ``Great, {\tool} uses the float window for guidance is a good idea.''; ``The guidance is much like the tutorial when the game software is first used. It can help us to understand a new app.''.
Participants also express that it can save their testing time such as ``Nice! The {\tool} saves our testing time!''.

\section{Conclusion}
\label{sec_conclusion}
As the last line of defence, manual testing is crucial to improve app quality.
Therefore, we develop the {\tool} to guide human testers in exploring more states during app testing.
We construct $STG_{action}$ and generate the planned path based on a DP algorithm.
On the app screen, we highlight the hint moves triggering the next unexplored state to users.
The automated evaluation and user study demonstrate the accuracy and usefulness of {\tool} in improving testing efficiency, reducing testing time and saving testers' efforts.
In the future, we will also improve the interaction between our {\tool} and users, which can borrow the idea from the human-machine collaboration studies to better facilitate the testers.


\begin{acks}
This work is supported by the National Key Research and Development Program of China under grant No.2018YFB1403400, National Natural Science Foundation
of China under Grant No. 62072442, No. 62002348, and Youth Innovation Promotion Association Chinese Academy of Sciences.
\end{acks}

\nocite{*} 

\bibliographystyle{ACM-Reference-Format}

\bibliography{reference}


\begin{thebibliography}{31}


\ifx \showCODEN    \undefined \def \showCODEN     #1{\unskip}     \fi
\ifx \showDOI      \undefined \def \showDOI       #1{#1}\fi
\ifx \showISBNx    \undefined \def \showISBNx     #1{\unskip}     \fi
\ifx \showISBNxiii \undefined \def \showISBNxiii  #1{\unskip}     \fi
\ifx \showISSN     \undefined \def \showISSN      #1{\unskip}     \fi
\ifx \showLCCN     \undefined \def \showLCCN      #1{\unskip}     \fi
\ifx \shownote     \undefined \def \shownote      #1{#1}          \fi
\ifx \showarticletitle \undefined \def \showarticletitle #1{#1}   \fi
\ifx \showURL      \undefined \def \showURL       {\relax}        \fi
\providecommand\bibfield[2]{#2}
\providecommand\bibinfo[2]{#2}
\providecommand\natexlab[1]{#1}
\providecommand\showeprint[2][]{arXiv:#2}

\bibitem[\protect\citeauthoryear{??}{Can}{2022}]%
        {Candy2020}
 \bibinfo{year}{2022}\natexlab{}.
\newblock \bibinfo{title}{Candy Crush Saga}.
\newblock \bibinfo{howpublished}{\url{https://www.king.com/game/candycrush}}.
\newblock


\bibitem[\protect\citeauthoryear{Alshayban, Ahmed, and Malek}{Alshayban
  et~al\mbox{.}}{2020}]%
        {alshayban2020accessibility}
\bibfield{author}{\bibinfo{person}{Abdulaziz Alshayban},
  \bibinfo{person}{Iftekhar Ahmed}, {and} \bibinfo{person}{Sam Malek}.}
  \bibinfo{year}{2020}\natexlab{}.
\newblock \showarticletitle{Accessibility issues in Android apps: state of
  affairs, sentiments, and ways forward}. In \bibinfo{booktitle}{\emph{ICSE}}.
\newblock


\bibitem[\protect\citeauthoryear{Chen, Su, Meng, Xing, and Liu}{Chen
  et~al\mbox{.}}{2018}]%
        {chen2018ui}
\bibfield{author}{\bibinfo{person}{Chunyang Chen}, \bibinfo{person}{Ting Su},
  \bibinfo{person}{Guozhu Meng}, \bibinfo{person}{Zhenchang Xing}, {and}
  \bibinfo{person}{Yang Liu}.} \bibinfo{year}{2018}\natexlab{}.
\newblock \showarticletitle{From ui design image to gui skeleton: a neural
  machine translator to bootstrap mobile gui implementation}. In
  \bibinfo{booktitle}{\emph{Proceedings of the 40th International Conference on
  Software Engineering}}. \bibinfo{pages}{665--676}.
\newblock


\bibitem[\protect\citeauthoryear{Chen, Chen, Hassan, Xing, Xia, and
  Hassan}{Chen et~al\mbox{.}}{2021b}]%
        {chen2021should}
\bibfield{author}{\bibinfo{person}{Qiuyuan Chen}, \bibinfo{person}{Chunyang
  Chen}, \bibinfo{person}{Safwat Hassan}, \bibinfo{person}{Zhengchang Xing},
  \bibinfo{person}{Xin Xia}, {and} \bibinfo{person}{Ahmed~E Hassan}.}
  \bibinfo{year}{2021}\natexlab{b}.
\newblock \showarticletitle{How Should I Improve the UI of My App? A Study of
  User Reviews of Popular Apps in the Google Play}.
\newblock \bibinfo{journal}{\emph{ACM Transactions on Software Engineering and
  Methodology (TOSEM)}} \bibinfo{volume}{30}, \bibinfo{number}{3}
  (\bibinfo{year}{2021}), \bibinfo{pages}{1--38}.
\newblock


\bibitem[\protect\citeauthoryear{Chen, Chen, Fan, Fan, Zhan, and Liu}{Chen
  et~al\mbox{.}}{2021a}]%
        {chen2021accessible}
\bibfield{author}{\bibinfo{person}{Sen Chen}, \bibinfo{person}{Chunyang Chen},
  \bibinfo{person}{Lingling Fan}, \bibinfo{person}{Mingming Fan},
  \bibinfo{person}{Xian Zhan}, {and} \bibinfo{person}{Yang Liu}.}
  \bibinfo{year}{2021}\natexlab{a}.
\newblock \showarticletitle{Accessible or Not An Empirical Investigation of
  Android App Accessibility}.
\newblock \bibinfo{journal}{\emph{IEEE Transactions on Software Engineering}}
  (\bibinfo{year}{2021}).
\newblock


\bibitem[\protect\citeauthoryear{Chen, Fan, Chen, Su, Li, Liu, and Xu}{Chen
  et~al\mbox{.}}{2019}]%
        {chen2019storydroid}
\bibfield{author}{\bibinfo{person}{Sen Chen}, \bibinfo{person}{Lingling Fan},
  \bibinfo{person}{Chunyang Chen}, \bibinfo{person}{Ting Su},
  \bibinfo{person}{Wenhe Li}, \bibinfo{person}{Yang Liu}, {and}
  \bibinfo{person}{Lihua Xu}.} \bibinfo{year}{2019}\natexlab{}.
\newblock \showarticletitle{Storydroid: Automated generation of storyboard for
  Android apps}. In \bibinfo{booktitle}{\emph{2019 IEEE/ACM 41st International
  Conference on Software Engineering (ICSE)}}. IEEE, \bibinfo{pages}{596--607}.
\newblock


\bibitem[\protect\citeauthoryear{Chen, Pandey, Song, and Oney}{Chen
  et~al\mbox{.}}{2020}]%
        {chen2020improving}
\bibfield{author}{\bibinfo{person}{Yan Chen}, \bibinfo{person}{Maulishree
  Pandey}, \bibinfo{person}{Song}, {and} \bibinfo{person}{Steve Oney}.}
  \bibinfo{year}{2020}\natexlab{}.
\newblock \showarticletitle{Improving Crowd-Supported GUI Testing with
  Structural Guidance}. In \bibinfo{booktitle}{\emph{CHI 2020}}.
\newblock


\bibitem[\protect\citeauthoryear{Feng, Ma, Yu, Chen, Zhou, and Zhen}{Feng
  et~al\mbox{.}}{2021}]%
        {feng2021auto}
\bibfield{author}{\bibinfo{person}{Sidong Feng}, \bibinfo{person}{Suyu Ma},
  \bibinfo{person}{Jinzhong Yu}, \bibinfo{person}{Chunyang Chen},
  \bibinfo{person}{TingTing Zhou}, {and} \bibinfo{person}{Yankun Zhen}.}
  \bibinfo{year}{2021}\natexlab{}.
\newblock \showarticletitle{Auto-icon: An automated code generation tool for
  icon designs assisting in ui development}. In \bibinfo{booktitle}{\emph{26th
  International Conference on Intelligent User Interfaces}}.
  \bibinfo{pages}{59--69}.
\newblock


\bibitem[\protect\citeauthoryear{Frisson, Malacria, Bailly, and Dutoit}{Frisson
  et~al\mbox{.}}{2016}]%
        {frisson2016inspectorwidget}
\bibfield{author}{\bibinfo{person}{Christian Frisson}, \bibinfo{person}{Sylvain
  Malacria}, \bibinfo{person}{Gilles Bailly}, {and} \bibinfo{person}{Thierry
  Dutoit}.} \bibinfo{year}{2016}\natexlab{}.
\newblock \showarticletitle{Inspectorwidget: A system to analyze users
  behaviors in their applications}. In \bibinfo{booktitle}{\emph{Proceedings of
  the 2016 CHI Conference Extended Abstracts on Human Factors in Computing
  Systems}}. \bibinfo{pages}{1548--1554}.
\newblock


\bibitem[\protect\citeauthoryear{Li, Yang, Guo, and Chen}{Li
  et~al\mbox{.}}{2017}]%
        {li2017droidbot}
\bibfield{author}{\bibinfo{person}{Yuanchun Li}, \bibinfo{person}{Ziyue Yang},
  \bibinfo{person}{Yao Guo}, {and} \bibinfo{person}{Xiangqun Chen}.}
  \bibinfo{year}{2017}\natexlab{}.
\newblock \showarticletitle{Droidbot: a lightweight ui-guided test input
  generator for android}. In \bibinfo{booktitle}{\emph{ICSE 2017}}.
\newblock


\bibitem[\protect\citeauthoryear{Li, Yang, Guo, and Chen}{Li
  et~al\mbox{.}}{2019}]%
        {li2019humanoid}
\bibfield{author}{\bibinfo{person}{Yuanchun Li}, \bibinfo{person}{Ziyue Yang},
  \bibinfo{person}{Yao Guo}, {and} \bibinfo{person}{Xiangqun Chen}.}
  \bibinfo{year}{2019}\natexlab{}.
\newblock \showarticletitle{Humanoid: a deep learning-based approach to
  automated black-box Android app testing}. In \bibinfo{booktitle}{\emph{2019
  34th IEEE/ACM International Conference on Automated Software Engineering
  (ASE)}}. IEEE, \bibinfo{pages}{1070--1073}.
\newblock


\bibitem[\protect\citeauthoryear{Linares-V{\'a}squez, Bernal-C{\'a}rdenas,
  Moran, and Poshyvanyk}{Linares-V{\'a}squez et~al\mbox{.}}{2017}]%
        {linares2017developers}
\bibfield{author}{\bibinfo{person}{Mario Linares-V{\'a}squez},
  \bibinfo{person}{Carlos Bernal-C{\'a}rdenas}, \bibinfo{person}{Kevin Moran},
  {and} \bibinfo{person}{Denys Poshyvanyk}.} \bibinfo{year}{2017}\natexlab{}.
\newblock \showarticletitle{How do developers test android applications?}. In
  \bibinfo{booktitle}{\emph{ICSME 2017}}.
\newblock


\bibitem[\protect\citeauthoryear{Liu, Chen, Wang, Huang, Hu, and Wang}{Liu
  et~al\mbox{.}}{2020}]%
        {liu2020owl}
\bibfield{author}{\bibinfo{person}{Zhe Liu}, \bibinfo{person}{Chunyang Chen},
  \bibinfo{person}{Junjie Wang}, \bibinfo{person}{Yuekai Huang},
  \bibinfo{person}{Jun Hu}, {and} \bibinfo{person}{Qing Wang}.}
  \bibinfo{year}{2020}\natexlab{}.
\newblock \showarticletitle{Owl Eyes: Spotting {UI} Display Issues via Visual
  Understanding}. In \bibinfo{booktitle}{\emph{35th {IEEE/ACM} International
  Conference on Automated Software Engineering, {ASE} 2020, Melbourne,
  Australia, September 21-25, 2020}}. \bibinfo{publisher}{{IEEE}},
  \bibinfo{pages}{398--409}.
\newblock
\urldef\tempurl%
\url{https://doi.org/10.1145/3324884.3416547}
\showDOI{\tempurl}


\bibitem[\protect\citeauthoryear{Liu, Chen, Wang, and Wang}{Liu
  et~al\mbox{.}}{2022a}]%
        {liu2022CHI}
\bibfield{author}{\bibinfo{person}{Zhe Liu}, \bibinfo{person}{Chunyang Chen},
  \bibinfo{person}{Junjie Wang}, {and} \bibinfo{person}{Qing Wang}.}
  \bibinfo{year}{2022}\natexlab{a}.
\newblock \showarticletitle{Guided Bug Crush: Assist Manual GUI Testing of
  Android Apps via Hint Moves}. In \bibinfo{booktitle}{\emph{CHI 2022}}.
\newblock
\urldef\tempurl%
\url{https://doi.org/10.1145/3491102.3501903}
\showDOI{\tempurl}


\bibitem[\protect\citeauthoryear{Liu, Chen, Wang, and Wang}{Liu
  et~al\mbox{.}}{2022b}]%
        {liu2022Nighthawk}
\bibfield{author}{\bibinfo{person}{Zhe Liu}, \bibinfo{person}{Chunyang Chen},
  \bibinfo{person}{Junjie Wang}, {and} \bibinfo{person}{Qing Wang}.}
  \bibinfo{year}{2022}\natexlab{b}.
\newblock \showarticletitle{Nighthawk: Fully Automated Localizing UI Display
  Issues via Visual Understanding}. In \bibinfo{booktitle}{\emph{IEEE
  Transactions on Software Engineering}}.
\newblock


\bibitem[\protect\citeauthoryear{Mao, Harman, and Jia}{Mao
  et~al\mbox{.}}{2016}]%
        {mao2016sapienz}
\bibfield{author}{\bibinfo{person}{Ke Mao}, \bibinfo{person}{Mark Harman},
  {and} \bibinfo{person}{Yue Jia}.} \bibinfo{year}{2016}\natexlab{}.
\newblock \showarticletitle{Sapienz: Multi-objective automated testing for
  Android applications}. In \bibinfo{booktitle}{\emph{Proceedings of the 25th
  International Symposium on Software Testing and Analysis}}.
  \bibinfo{pages}{94--105}.
\newblock


\bibitem[\protect\citeauthoryear{Pan, Huang, Wang, Zhang, and Li}{Pan
  et~al\mbox{.}}{2020}]%
        {pan2020reinforcement}
\bibfield{author}{\bibinfo{person}{Minxue Pan}, \bibinfo{person}{An Huang},
  \bibinfo{person}{Guoxin Wang}, \bibinfo{person}{Tian Zhang}, {and}
  \bibinfo{person}{Xuandong Li}.} \bibinfo{year}{2020}\natexlab{}.
\newblock \showarticletitle{Reinforcement learning based curiosity-driven
  testing of Android apps}. In \bibinfo{booktitle}{\emph{ISSTA}}.
\newblock


\bibitem[\protect\citeauthoryear{Su, Liu, Chen, Wang, and Wang}{Su
  et~al\mbox{.}}{2021}]%
        {DBLP:conf/sigsoft/SuLC0W21}
\bibfield{author}{\bibinfo{person}{Yuhui Su}, \bibinfo{person}{Zhe Liu},
  \bibinfo{person}{Chunyang Chen}, \bibinfo{person}{Junjie Wang}, {and}
  \bibinfo{person}{Qing Wang}.} \bibinfo{year}{2021}\natexlab{}.
\newblock \showarticletitle{OwlEyes-online: a fully automated platform for
  detecting and localizing {UI} display issues}. In
  \bibinfo{booktitle}{\emph{{ESEC/FSE} '21: 29th {ACM} Joint European Software
  Engineering Conference and Symposium on the Foundations of Software
  Engineering, Athens, Greece, August 23-28, 2021}}.
  \bibinfo{publisher}{{ACM}}, \bibinfo{pages}{1500--1504}.
\newblock
\urldef\tempurl%
\url{https://doi.org/10.1145/3468264.3473109}
\showDOI{\tempurl}


\bibitem[\protect\citeauthoryear{Wang, Wang, Chen, Menzies, Cui, Xie, and
  Wang}{Wang et~al\mbox{.}}{2021a}]%
        {DBLP:journals/tse/WangWCMCXW21}
\bibfield{author}{\bibinfo{person}{Junjie Wang}, \bibinfo{person}{Song Wang},
  \bibinfo{person}{Jianfeng Chen}, \bibinfo{person}{Tim Menzies},
  \bibinfo{person}{Qiang Cui}, \bibinfo{person}{Miao Xie}, {and}
  \bibinfo{person}{Qing Wang}.} \bibinfo{year}{2021}\natexlab{a}.
\newblock \showarticletitle{Characterizing Crowds to Better Optimize Worker
  Recommendation in Crowdsourced Testing}.
\newblock \bibinfo{journal}{\emph{{IEEE} Trans. Software Eng.}}
  \bibinfo{volume}{47}, \bibinfo{number}{6} (\bibinfo{year}{2021}),
  \bibinfo{pages}{1259--1276}.
\newblock
\urldef\tempurl%
\url{https://doi.org/10.1109/TSE.2019.2918520}
\showDOI{\tempurl}


\bibitem[\protect\citeauthoryear{Wang, Yang, Krishna, Menzies, and Wang}{Wang
  et~al\mbox{.}}{2019}]%
        {wang2019iSENSE}
\bibfield{author}{\bibinfo{person}{Junjie Wang}, \bibinfo{person}{Ye Yang},
  \bibinfo{person}{Rahul Krishna}, \bibinfo{person}{Tim Menzies}, {and}
  \bibinfo{person}{Qing Wang}.} \bibinfo{year}{2019}\natexlab{}.
\newblock \showarticletitle{iSENSE: Completion-Aware Crowdtesting Management}.
  In \bibinfo{booktitle}{\emph{ICSE'2019}}. \bibinfo{pages}{932--943}.
\newblock


\bibitem[\protect\citeauthoryear{Wang, Yang, Wang, Chen, Wang, and Wang}{Wang
  et~al\mbox{.}}{2021b}]%
        {wang2021context}
\bibfield{author}{\bibinfo{person}{Junjie Wang}, \bibinfo{person}{Ye Yang},
  \bibinfo{person}{Song Wang}, \bibinfo{person}{Chunyang Chen},
  \bibinfo{person}{Dandan Wang}, {and} \bibinfo{person}{Qing Wang}.}
  \bibinfo{year}{2021}\natexlab{b}.
\newblock \showarticletitle{Context-aware Personalized Crowdtesting Task
  Recommendation}.
\newblock \bibinfo{journal}{\emph{IEEE Transactions on Software Engineering}}
  (\bibinfo{year}{2021}).
\newblock


\bibitem[\protect\citeauthoryear{Wang, Yang, Wang, Hu, Wang, and Wang}{Wang
  et~al\mbox{.}}{2020}]%
        {wang2020context}
\bibfield{author}{\bibinfo{person}{Junjie Wang}, \bibinfo{person}{Ye Yang},
  \bibinfo{person}{Song Wang}, \bibinfo{person}{Yuanzhe Hu},
  \bibinfo{person}{Dandan Wang}, {and} \bibinfo{person}{Qing Wang}.}
  \bibinfo{year}{2020}\natexlab{}.
\newblock \showarticletitle{Context-aware In-process Crowdworker
  Recommendation} \emph{(\bibinfo{series}{ICSE 2020})}.
\newblock


\bibitem[\protect\citeauthoryear{Wang, Yang, Xu, and Xie}{Wang
  et~al\mbox{.}}{2021c}]%
        {Wang2021Vet}
\bibfield{author}{\bibinfo{person}{Wenyu Wang}, \bibinfo{person}{Wei Yang},
  \bibinfo{person}{Tianyin Xu}, {and} \bibinfo{person}{Tao Xie}.}
  \bibinfo{year}{2021}\natexlab{c}.
\newblock \showarticletitle{Vet: Identifying and Avoiding UI Exploration
  Tarpits}. In \bibinfo{booktitle}{\emph{ESEC/FSE 2021}}.
\newblock


\bibitem[\protect\citeauthoryear{Xie, Feng, Xing, Chen, and Chen}{Xie
  et~al\mbox{.}}{2020}]%
        {DBLP:conf/sigsoft/XieFXCC20}
\bibfield{author}{\bibinfo{person}{Mulong Xie}, \bibinfo{person}{Sidong Feng},
  \bibinfo{person}{Zhenchang Xing}, \bibinfo{person}{Jieshan Chen}, {and}
  \bibinfo{person}{Chunyang Chen}.} \bibinfo{year}{2020}\natexlab{}.
\newblock \showarticletitle{{UIED:} a hybrid tool for {GUI} element detection}.
  In \bibinfo{booktitle}{\emph{{ESEC/FSE} '20: 28th {ACM} Joint European
  Software Engineering Conference and Symposium on the Foundations of Software
  Engineering, Virtual Event, USA, November 8-13, 2020}},
  \bibfield{editor}{\bibinfo{person}{Prem Devanbu}, \bibinfo{person}{Myra~B.
  Cohen}, {and} \bibinfo{person}{Thomas Zimmermann}} (Eds.).
  \bibinfo{publisher}{{ACM}}, \bibinfo{pages}{1655--1659}.
\newblock
\urldef\tempurl%
\url{https://doi.org/10.1145/3368089.3417940}
\showDOI{\tempurl}


\bibitem[\protect\citeauthoryear{Yan, Pan, Yan, and Liang}{Yan
  et~al\mbox{.}}{2020}]%
        {yanmultiple}
\bibfield{author}{\bibinfo{person}{Jiwei Yan}, \bibinfo{person}{Linjie Pan},
  \bibinfo{person}{Jun Yan}, {and} \bibinfo{person}{Bin Liang}.}
  \bibinfo{year}{2020}\natexlab{}.
\newblock \showarticletitle{Multiple-entry testing of android applications by
  constructing activity launching contexts}. In \bibinfo{booktitle}{\emph{ICSE
  2020}}.
\newblock


\bibitem[\protect\citeauthoryear{Yang, Xing, Xia, Chen, Ye, and Li}{Yang
  et~al\mbox{.}}{2021a}]%
        {yang2021don}
\bibfield{author}{\bibinfo{person}{Bo Yang}, \bibinfo{person}{Zhenchang Xing},
  \bibinfo{person}{Xin Xia}, \bibinfo{person}{Chunyang Chen},
  \bibinfo{person}{Deheng Ye}, {and} \bibinfo{person}{Shanping Li}.}
  \bibinfo{year}{2021}\natexlab{a}.
\newblock \showarticletitle{Don’t Do That! Hunting Down Visual Design Smells
  in Complex UIs against Design Guidelines}. In \bibinfo{booktitle}{\emph{2021
  IEEE/ACM 43rd International Conference on Software Engineering (ICSE)}}.
  IEEE, \bibinfo{pages}{761--772}.
\newblock


\bibitem[\protect\citeauthoryear{Yang, Xing, Xia, Chen, Ye, and Li}{Yang
  et~al\mbox{.}}{2021b}]%
        {yang2021uis}
\bibfield{author}{\bibinfo{person}{Bo Yang}, \bibinfo{person}{Zhenchang Xing},
  \bibinfo{person}{Xin Xia}, \bibinfo{person}{Chunyang Chen},
  \bibinfo{person}{Deheng Ye}, {and} \bibinfo{person}{Shanping Li}.}
  \bibinfo{year}{2021}\natexlab{b}.
\newblock \showarticletitle{UIS-Hunter: Detecting UI Design Smells in Android
  Apps}. In \bibinfo{booktitle}{\emph{2021 IEEE/ACM 43rd International
  Conference on Software Engineering: Companion Proceedings (ICSE-Companion)}}.
  IEEE, \bibinfo{pages}{89--92}.
\newblock


\bibitem[\protect\citeauthoryear{Yang, Wu, Zhang, Wang, Swaminathan, Yan, and
  Rountev}{Yang et~al\mbox{.}}{2018}]%
        {yang2018static}
\bibfield{author}{\bibinfo{person}{Shengqian Yang}, \bibinfo{person}{Haowei
  Wu}, \bibinfo{person}{Hailong Zhang}, \bibinfo{person}{Yan Wang},
  \bibinfo{person}{Chandrasekar Swaminathan}, \bibinfo{person}{Dacong Yan},
  {and} \bibinfo{person}{Atanas Rountev}.} \bibinfo{year}{2018}\natexlab{}.
\newblock \showarticletitle{Static window transition graphs for Android}.
\newblock \bibinfo{journal}{\emph{Automated Software Engineering}}
  \bibinfo{volume}{25}, \bibinfo{number}{4} (\bibinfo{year}{2018}),
  \bibinfo{pages}{833--873}.
\newblock


\bibitem[\protect\citeauthoryear{Zhang, Sui, and Xue}{Zhang
  et~al\mbox{.}}{2018}]%
        {zhang2018launch}
\bibfield{author}{\bibinfo{person}{Yifei Zhang}, \bibinfo{person}{Yulei Sui},
  {and} \bibinfo{person}{Jingling Xue}.} \bibinfo{year}{2018}\natexlab{}.
\newblock \showarticletitle{Launch-mode-aware context-sensitive activity
  transition analysis}. In \bibinfo{booktitle}{\emph{ICSE 2018}}.
\newblock


\bibitem[\protect\citeauthoryear{Zhao, Xing, Chen, Xu, Zhu, Li, and Wang}{Zhao
  et~al\mbox{.}}{2020}]%
        {zhao2020seenomaly}
\bibfield{author}{\bibinfo{person}{Dehai Zhao}, \bibinfo{person}{Zhenchang
  Xing}, \bibinfo{person}{Chunyang Chen}, \bibinfo{person}{Xiwei Xu},
  \bibinfo{person}{Liming Zhu}, \bibinfo{person}{Guoqiang Li}, {and}
  \bibinfo{person}{Jinshui Wang}.} \bibinfo{year}{2020}\natexlab{}.
\newblock \showarticletitle{Seenomaly: vision-based linting of GUI animation
  effects against design-don't guidelines}. In \bibinfo{booktitle}{\emph{2020
  IEEE/ACM 42nd International Conference on Software Engineering (ICSE)}}.
  IEEE, \bibinfo{pages}{1286--1297}.
\newblock


\bibitem[\protect\citeauthoryear{Zhao, Chen, Liu, and Zhu}{Zhao
  et~al\mbox{.}}{2021}]%
        {zhao2021guigan}
\bibfield{author}{\bibinfo{person}{Tianming Zhao}, \bibinfo{person}{Chunyang
  Chen}, \bibinfo{person}{Yuanning Liu}, {and} \bibinfo{person}{Xiaodong Zhu}.}
  \bibinfo{year}{2021}\natexlab{}.
\newblock \showarticletitle{GUIGAN: Learning to Generate GUI Designs Using
  Generative Adversarial Networks}. In \bibinfo{booktitle}{\emph{2021 IEEE/ACM
  43rd International Conference on Software Engineering (ICSE)}}. IEEE,
  \bibinfo{pages}{748--760}.
\newblock


\end{thebibliography}

\end{document}